\documentclass[onecolumn]{aa}
\usepackage[varg]{txfonts}
\usepackage{graphicx}
\usepackage{natbib}
\usepackage{CJK}

\begin{document}
\begin{CJK}{UTF8}{bkai}   
                          
\title{Six Faint Gamma-ray Pulsars Seen with the Fermi Large Area Telescope}
\subtitle{Towards a Sample Blending into the Background}

\author{
X.~Hou(侯賢)$^{(1)}$ \and 
D.~A.~Smith$^{(1)}$ \and 
L.~Guillemot$^{(2)}$ \and 
C.~C.~Cheung$^{(3)}$ \and 
I.~Cognard$^{(2)}$ \and
H.~A.~Craig$^{(4)}$ \and 
C.~M.~Espinoza$^{(5)}$ \and 
S.~Johnston$^{(6)}$ \and 
M.~Kramer$^{(7,8)}$ \and 
O.~Reimer$^{(9,4)}$ \and 
T.~Reposeur$^{(1)}$ \and 
R.~Shannon$^{(6)}$ \and 
B.~W.~Stappers$^{(7)}$ \and
P.~Weltevrede$^{(7)}$
}

\institute{
\inst{1}~Centre d'\'Etudes Nucl\'eaires de Bordeaux Gradignan, IN2P3/CNRS, Universit\'e Bordeaux 1, BP120, F-33175 Gradignan Cedex, France\\ 
\inst{2}~Laboratoire de Physique et Chimie de l'Environnement et de l'Espace -- Universit\'e d'Orl\'eans / CNRS, F-45071 Orl\'eans Cedex 02, France and Station de radioastronomie de Nan\c{c}ay, Observatoire de Paris, CNRS/INSU, F-18330 Nan\c{c}ay, France\\ 
\inst{3}~Space Science Division, Naval Research Laboratory, Washington, DC 20375-5352, USA\\ 
\inst{4}~W. W. Hansen Experimental Physics Laboratory, Kavli Institute for Particle Astrophysics and Cosmology, Department of Physics and SLAC National Accelerator Laboratory, Stanford University, Stanford, CA 94305, USA\\ 
\inst{5}~Instituto de Astrof\'isica, Facultad de F\'isica, Pontificia Universidad Cat\'olica de Chile, Casilla 306, Santiago 22, Chile\\
\inst{6}~CSIRO Astronomy and Space Science, Australia Telescope National Facility, Epping NSW 1710, Australia\\ 
\inst{7}~Jodrell Bank Centre for Astrophysics, School of Physics and Astronomy, The University of Manchester, M13 9PL, UK\\ 
\inst{8}~Max-Planck-Institut f\"ur Radioastronomie, Auf dem H\"ugel 69, 53121 Bonn, Germany\\ 
\inst{9}~Institut f\"ur Astro- und Teilchenphysik and Institut f\"ur Theoretische Physik, Leopold-Franzens-Universit\"at Innsbruck, A-6020 Innsbruck, Austria\\ 
\email{smith@cenbg.in2p3.fr} \\
\email{xianhou.astro@gmail.com} \\
\email{lucas.guillemot@cnrs-orleans.fr} \\
}

\date{Received: \hspace*{3cm}; accepted: }

\authorrunning{Hou, Smith et al.}
\titlerunning{Six faint gamma-ray pulsars}

\abstract
{
GeV gamma-ray pulsations from over 140 pulsars have been characterized using the \textit{Fermi} Large Area Telescope, 
enabling improved understanding of the emission regions within the neutron star magnetospheres, 
and the contributions of pulsars to high energy electrons and diffuse gamma rays in the Milky Way. 
The first gamma-ray pulsars to be detected were the most intense and/or those with narrow pulses.
 }
{
As the \textit{Fermi} mission progresses, progressively fainter objects can be studied. 
In addition to more distant pulsars (thus probing a larger volume of the Galaxy), or ones in high background regions
(thus improving the sampling uniformity across the Galactic plane),
we detect pulsars with broader pulses or lower luminosity. 
Adding pulsars to our catalog with inclination angles that are rare in the observed sample, 
and/or with lower spindown power, will reduce the bias in the currently known gamma-ray pulsar population.
}
{
We use rotation ephemerides derived from radio observations to phase-fold gamma rays recorded by the \textit{Fermi} Large Area Telescope, 
to then determine the pulse profile properties. Spectral analysis provides the luminosities and, when the signal-to-noise ratio allows, the cutoff energies.
We constrain the pulsar distances by different means in order to minimize the luminosity uncertainties.
}
{
We present six new gamma-ray pulsars with an eclectic mix of properties. Three are young, and three are recycled. 
They include the farthest, the lowest power, two of the highest duty-cycle pulsars seen, 
and only the fourth young gamma-ray pulsar with a radio interpulse. We discuss the biases existing in the current
gamma-ray pulsar catalog, and steps to be taken to mitigate the bias.
}
{}

\keywords{observations -- pulsars: general -- pulsars: individual (J1055$-$6028, J1640+2224, J1705$-$1906, J1732$-$5049, J1843$-$1113, J1913+0904)-- stars: neutron -- stars:
individual: AG Car, GG Car}   

\maketitle

\section{Introduction}
Excellent as it may be, the ``Second \textit{Fermi} LAT Pulsar Catalog'' \citep[][hereafter 2PC]{2PC} is biased.
At launch, and early in the mission of NASA's \textit{Fermi Gamma-ray Space Telescope}, 
a primary goal for the Large Area Telescope (LAT) instrument team and the Pulsar Timing Consortium \citep[PTC, ][]{TimingForFermi}
was the unambiguous identification and characterization of gamma-ray pulsars. 
We thus applied strict criteria to identify gamma-ray emission that is highly modulated over a neutron star rotation.
These criteria favor sources bright enough to identify the spectral cutoffs typical of gamma-ray pulsars. 

\textit{Fermi} has begun the second half of its nominal ten-year mission, and our focus is shifting to a more
uniform sampling of the gamma-ray pulsar population. Gamma-ray emission models \citep[e.g.][]{AtlasII} predict that
for some magnetic inclinations, $\alpha$, and viewing angles, $\zeta$, the emission changes little
with rotational phase, and thus pulsations become more difficult to detect.
Simple geometry suggests that these $(\alpha, \, \zeta)$ combinations are relatively rare: to validate the models via a few detections, 
a large neutron star population must be sampled. This means probing a large space volume, i.e., searching to larger distances, $d$. 
For Galactic pulsars with some distribution of heights above and below the plane, larger distances imply, overall, lower Galactic latitudes, 
and thus higher background levels due to the bright diffuse emission near the Galactic plane.
To compound the problem, 
predicted modulation weakens for pulsar spindown powers $\dot E$ approaching the empirical ``deathline'' near $3\times 10^{33}$ erg s$^{-1}$ discussed in 2PC, 
where gamma-ray emission seems to vanish. 
Since the luminosity $L_\gamma$ scales with some low power of $\dot E$, the flux scales as $L_\gamma/d^2$, and detectability
of a steady source depends on the flux-to-background level, an expanded gamma-ray pulsar sample will include some very faint
detections, blending into the background at times. 

The eclectic mix of six new gamma-ray pulsars presented here have in common that they are all faint in one way or another,
where ``faint'' in this paper means difficult to detect or characterize. 
Table \ref{BigTab} summarizes their properties.
They were discovered during the ongoing program of routinely gamma-ray phase-folding as many known pulsars as possible, 
using the over 800 rotation ephemerides provided by the PTC.
They were mentioned in 2PC, with some detail provided by \citet{Hou2013}.
The present work goes into more depth, especially for their spectra.
They illustrate the challenges that arise in trying to compare observed pulsar samples with the predictions of population syntheses
as weaker and weaker sources become accessible to the LAT.
They may thus help define future strategies to make the observed sample as complete as possible.

\section{Observations and Analysis}
\subsection{Gamma-ray data}

The LAT is an electron-positron pair conversion telescope on \textit{Fermi},  launched on 2008 June 11 \citep{LATinstrument}. 
Sensitive to gamma rays in the 20 MeV to $>300$ GeV energy band, the LAT has better sensitivity and localization than previous instruments. 
With the reprocessed {\it Pass 7} data used for the analyses presented here, 
the LAT \emph{Instrument Response Functions} (IRFs) have an on-axis effective area of $\sim 7000$ cm$^2$ above 1\,GeV (P7REP\_V15).
Below several GeV multiple scattering dominates the LAT's angular resolution.
The on-axis instrument \emph{point-spread function} (PSF) provides 68\% confidence regions for gamma-ray direction reconstruction within 
angular radii slightly more than $5^\circ$ at the lowest energies used here ($0.1$ GeV),
decreasing to under $0.4^\circ$ at 3 GeV, at which energy the spectral rollover has become pronounced for most pulsars.

Table \ref{DataTab} details the gamma-ray dataset, background models, and standard \textit{Science Tools} analysis software we used\footnote{Gamma-ray data, 
analysis software, rotation ephemerides, and the diffuse background models are publically available 
at the \textit{Fermi} Science Support Center, FSSC, http://fermi.gsfc.nasa.gov/ssc/.}. 
We used over 52 months of P7REP data, keeping ``source'' class events (high probability of being gamma-ray photons) with energies between $0.1$ and 100 GeV 
within a $10^\circ$ radius ``region-of-interest'' (ROI). 
The center of the ROI is offset from the pulsar position, explained in Section 2.3.
The broad PSF imposes a large region for the spectral analysis, reduced for the pulse profile study, as described below.
The zenith angle cut ($<100^\circ$) rejects atmospheric gamma rays from the Earth's limb.
We kept only events that had good quality flags and were
collected when the LAT rocking angle (angle between the normal to the LAT's front surface and the orbital plane) was smaller than $52^{\circ}$. 
Observation times when a pulsar was within $10^\circ$ of the Sun or Moon's direction were excluded, to remove possible contamination by gamma rays created
when cosmic rays graze their surfaces\footnote{The Sun's position is included in the LAT spacecraft ``FT2'' data files. The moon's can be added to the FT2
using \emph{``moonpos''} available at http://fermi.gsfc.nasa.gov/ssc/data/analysis/user/}. 
This does not affect the two pulsars far from the ecliptic plane, PSRs J1055$-$6028 and J1732$-$5049.

\begin{table*}[ht]
\caption{\small Properties of the three young gamma-ray pulsars (top) and three gamma-ray MSPs (bottom). 
$\dot E$ and $B_{\rm LC}$ for pulsars with proper motion measurements are Doppler corrected as in 2PC,
with uncertainties from the proper motions and distances.
The first spectral uncertainties are statistical, the second are systematic. 
The luminosity assumes a beaming factor $f_{\Omega}=1$. 
Profile fit types -- G: Gaussian; G2: 2-sided Gaussian; L: Lorentzian; L2: 2-sided Lorentzian.
}
\begin{tabular}{lccc}
Pulsar name   & J1055$-$6028\tablefootmark{a} &J1705$-$1906 &J1913+0904  \\
Galactic longitude, latitude ($l,b$)  &$(289.13^\circ,\, -0.75^\circ)$ &$(3.19^\circ,\, 13.03^\circ)$ &$(43.50^\circ,\, -0.68^\circ)$  \\
Spin period, $P$ (ms)  & 99.7 &299.0  &163.2 \\
Dispersion measure, DM ($\rm pc\, cm^{-3}$)  &$635.9$ &$22.91$ &$97.27$  \\
Flux density at 1.4 GHz, $S_{14}$ (mJy) & 0.78 & 8 & 0.07 \\
Spin-down power, $\dot E$ ($10^{33}\rm erg\,s^{-1}$)  & 1180 &6.11 &160 \\
Characteristic age, $\tau$ (kyr)  &53.5 & 1140 & 147 \\
Field at light cylinder, $B_{\rm LC}$ (kG)  & 16.4 & 0.4 & 3.7 \\
Distance, $d$ (kpc)  &$8^{+5}_{-3}$ &$0.9\pm0.1$ &$3.0\pm0.4$ \\
Observatory  & PKS & PKS &JBO \\
$N_{\rm TOA}$  &102 &73 &195 \\
Timing residual RMS ($\mu$s)  & 981 & 248 &1400 \\
Ephemeris validity range (MJD)   &$54505-56397$ &$54220-56397$ &$54588-56423$ \\ 
ROI centers $(l,b)$  & $(293.0^\circ,\, -1.8^\circ)$  & $(2.1^\circ,\, 15.1^\circ)$ & $(46.7^\circ,\, -0.6^\circ)$ \\ 
TS  (TS$_{\rm cut}$) &334 (54)&31 (12)&139 (40)\\
Spectral index, $\Gamma$    &$1.6 \pm 0.2 \pm 0.1$ & $2.3 \pm 0.2 \pm 0.2$ & $1.5 \pm 0.3\pm 0.6 $ \\
Cutoff energy, $E_{\rm c}$ (GeV)  &$2.2 \pm 0.5 \pm 0.7$ &   ...  &$1.6 \pm 0.8 \pm 1.3$ \\
Integral energy flux, $G_{\rm 100}$ ($10^{-12}\,\rm erg\,cm^{-2}\,s^{-1}$)  &$36 \pm 4 \pm 8$ &$2.7 \pm 0.8 \pm 0.2 $ &$31 \pm 5 \pm 14$ \\
Luminosity, $L_\gamma\, (10^{33}\,\rm erg\,s^{-1})$  & $280 \pm 30^{{+460}}_{{-180}} $ &$0.25 \pm 0.08\pm 0.06$ &$34 \pm 5 \pm 18$ \\
Efficiency, $\eta$ (\%) & $24 \pm 3^{{+40}}_{{-15}} $&$4.0 \pm 1.3\pm 1.0$ & $21 \pm 3\pm 11$  \\
Weighted $H$-test (significance)  &61 ($6.7\sigma$) &39 ($5.2\sigma$)& 54 ($6.2\sigma$)\\
$N_{\rm peak}$  &2(3) &1 &2(3) \\
Radio lag, $\delta$ (phase) &$0.13\pm0.05$ &$0.57\pm0.01$ &$0.33\pm0.04$ \\
$\gamma$-ray peak separation, $\Delta$ (phase) &$0.31\pm0.05$ &... &$0.32\pm0.04$ \\
On-peak definition (phase)  &$0.90-0.70$ &$0.40-0.65$ &$0.0-0.8$  \\
Lightcurve fit type   &G  &L  &G \\
 \\ \hline \\
Pulsar name   & J1640+2224   & J1732$-$5049 &J1843$-$1113   \\
Galactic longitude, latitude ($l,b$)   &$(41.05^\circ,\, 38.27^\circ)$  &$(340.03^\circ,\, -9.45^\circ)$ &$(22.06^\circ,\, -3.40^\circ)$  \\
Spin period, $P$ (ms) &3.16 & 5.31  &1.85   \\
Dispersion measure, DM ($\rm pc\, cm^{-3}$) &$18.43$ &$56.83$    &$59.97$   \\
Flux density at 1.4 GHz, $S_{14}$ (mJy) & 2 & 1.7 & 0.10 \\
Binary period, $P_b$ (days) & 175.4607 &5.2630 & ... \\
Proper motion in right ascension, $\mu_{\alpha} \cos({\rm dec})$ ($\rm mas\; yr^{-1}$) & $2.10\pm 0.03$ & $-0.51\pm0.11$  & $-2.17\pm0.07$  \\
Proper motion in declination, $\mu_{\rm dec}$ ($\rm mas\; yr^{-1}$) & $-11.2\pm 0.07$ &$-9.90\pm0.22$ &$-2.74\pm0.25$  \\
Timing parallax, (mas) & $2.2 \pm 1.3$  & ... & ...\\
Spin-down power, $\dot E$ ($10^{33}\rm erg\,s^{-1}$) &$3.3^{+0.8}_{-0.2} $ & $3.2\pm0.1$ &$57.8\pm0.02$  \\
Field at light cylinder, $B_{\rm LC}$ ($10^3$ G) & $27.0^{+3.5}_{-0.9}$ & $15.8 \pm 0.2$ & $192.7 \pm 0.3$  \\
Distance, $d$ (kpc) &$0.45^{+0.66}_{-0.16}$ &$1.4\pm0.2$ &$1.7\pm0.2$ \\
Observatory & NRT & PKS & NRT  \\
$N_{\rm TOA}$ & 154 &344 &107 \\
Timing residual RMS ($\mu$s) & 0.96 & 3.3 & 0.95  \\
Ephemeris validity range (MJD) &$53312-56438$  &$52647-55724$ &$55040-56473$  \\
ROI centers $(l,b)$  & Pulsar position  & $(340.1^\circ,\, -11.3^\circ)$ & $(25.0^\circ,\, -8.4^\circ)$ \\ 
TS (TS$_{\rm cut}$)&32 (1.6)&65 (7)&83 (6) \\
Spectral index, $\Gamma$   &$2.1 \pm 0.2 \pm 0.1 $ &$2.2 \pm 0.1 \pm 0.1 $ &$2.8 \pm 0.1 \pm0.2$  \\
Integral energy flux, $G_{\rm 100}$ ($10^{-12}\,\rm erg\,cm^{-2}\,s^{-1}$) &$2.3 \pm 0.6 \pm 0.2$ &$7.0 \pm 1.1 \pm 1.4$ & $15.5 \pm 2.1^{{+2.1}}_{{-7.3}} $ \\
Luminosity, $L_\gamma\, (10^{33}\,\rm erg\,s^{-1})$  & $0.056 \pm 0.013^{{+0.28}}_{{-0.03}} $ &$1.7 \pm 0.3\pm 0.6$ & $5.4 \pm 0.7^{{+0.7}}_{{-2.5}} $ \\
Efficiency, $\eta$ (\%)& $1.6 \pm 0.4^{{+8.0}}_{{-1.0}} $ &$52 \pm 8^{{+20}}_{{-17}} $ &$ 9.0 \pm 1.2^{{+1.2}}_{{-4.0}} $  \\ 
Weighted $H$-test (significance) &45 ($5.7\sigma$)&36 ($5.0\sigma$)&49 ($5.9\sigma$) \\
$N_{\rm peak}$ &1 &2 &1  \\
Radio lag, $\delta$ (phase) &$0.48\pm0.03$ &$0.39\pm0.04$ &$0.09\pm0.01$  \\
$\gamma$-ray peak separation, $\Delta$ (phase) &... &$0.27\pm0.04$ & ... \\
On-peak definition (phase) &$0.21-0.61$ &$0.25-0.80$ &$0.80-0.20$  \\
Lightcurve fit type &G  &G2  &L2  \\
\label{BigTab}
\end{tabular}
\tablefoot{\tablefoottext{a}{At discovery was misnamed J1055$-$6032.}}
\end{table*}

\clearpage

\subsection{Radio Timing and Gamma-ray Profiles}
Ground tests of the satellite GPS-based clock system, followed by on-orbit monitoring, demonstrated an absolute event timestamp precision of $<1\, \mu$s relative to UTC \citep{OnOrbit}. 
The PTC regularly times over 800 radio pulsars, providing rotation ephemerides with which to phase-fold gamma-ray photons \citep{TimingForFermi}.
This has been a great success, allowing detections of more than half of the gamma-ray pulsars to date. 
The others were discovered in blind period searches of the gamma-ray data, see for example \citet{Pletsch_2012}. 
Table \ref{BigTab} lists the radio observatories and the validity date ranges for the ephemerides used here. 
The radio data used in this work came from the Lovell telescope at Jodrell Bank Observatory \citep[JBO,][]{Jodrell};
from the Parkes observatory \citep[PKS,][]{ParkesFermiTiming}; and from the Nan\c cay radio telescope \citep[NRT,][]{Cognard2011}.
The ephemerides completely cover the gamma-ray data sample, except for the millisecond pulsar (MSP) PSR J1732$-$5049. 
We extrapolated this highly stable pulsar's timing solution to the end of the data, and found that the pulsed significance increased uniformly.

Timing of PSR J1640+2224 using 154 observations acquired over nine years with the Nan\c cay radiotelescope yields a parallax of $\pi = 2.2 \pm 1.3$ mas.
The root-mean-square of the timing residuals, weighted for the uncertainties on each measurement, is just under $1\, \mu$s. 
\citet{Lohmer2005} previously constrained its timing parallax to $\pi < 3.7$ mas.
PSR J1843$-$1113 also has sub-microsecond residuals, slightly more than the expected parallax shift at the lower end of the distance
range derived from the Dispersion Measure (DM, the electron column density along the line of sight measured during pulsar observations at radio frequencies).
However, we detect no parallax for this pulsar. Distances are discussed in Section 2.4.

\begin{figure*}[h]
\centering
\includegraphics[width=0.33\textwidth]{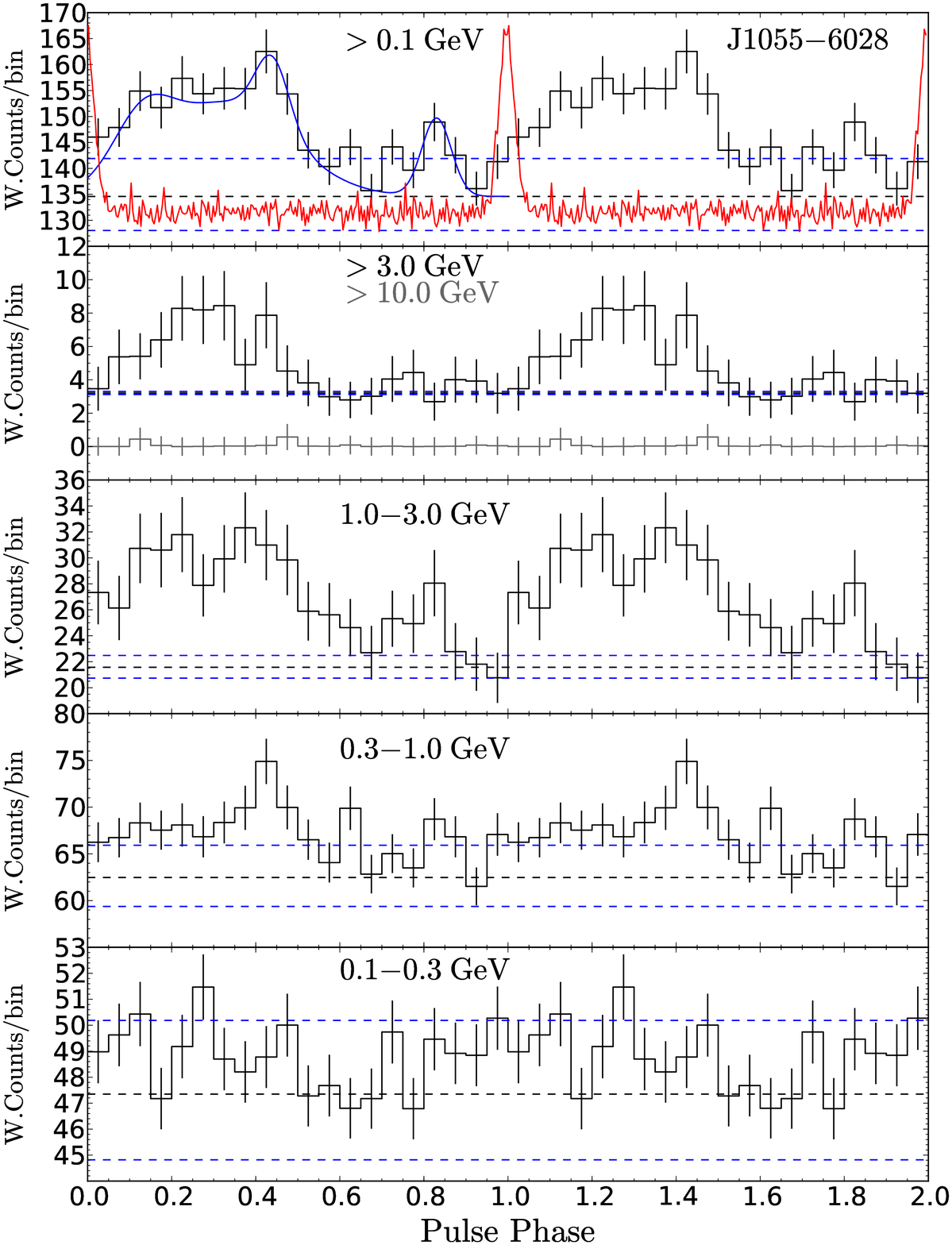}
\includegraphics[width=0.33\textwidth]{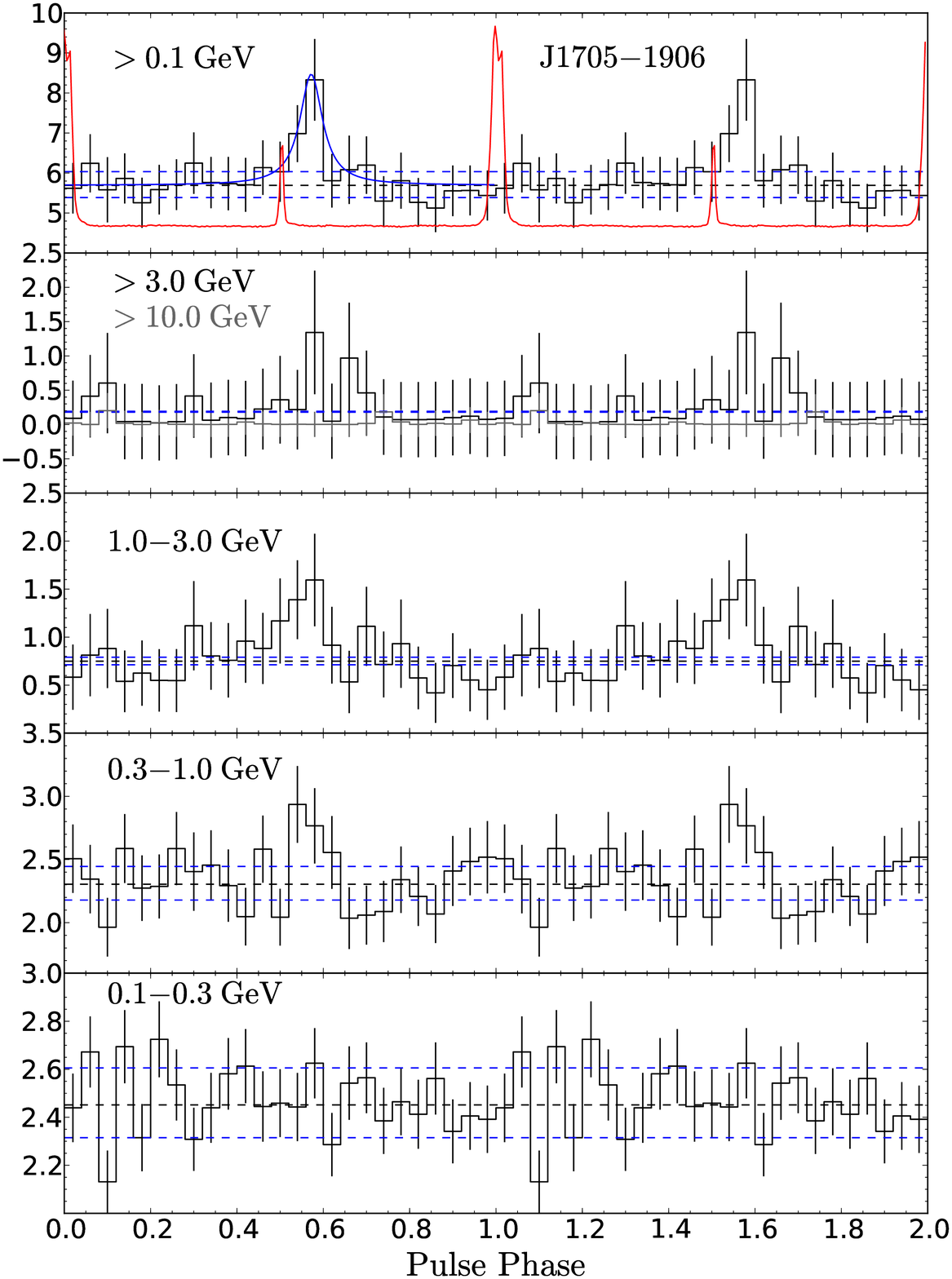}
\includegraphics[width=0.33\textwidth]{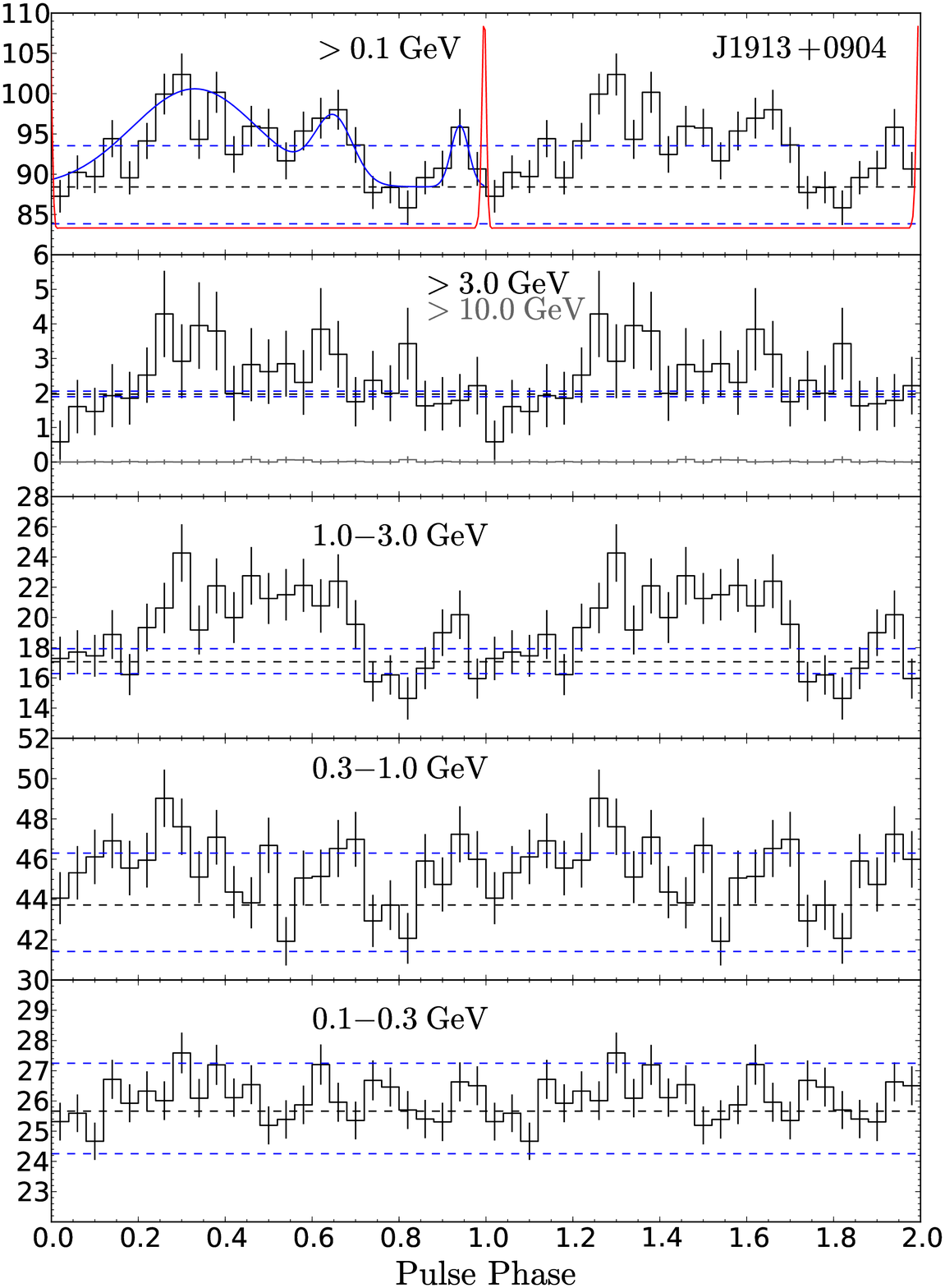}
\caption{Gamma-ray pulse profiles (black histograms) for the young pulsars PSR J1055$-$6028 (left), J1705$-$1906 (middle), and J1913+0904 (right). 
Each gamma-ray event is weighted, and the error bars are explained in Section 2.2.
The horizontal dotted lines are the background levels, with $\pm 1\sigma$ uncertainties.
Each pulsar rotation is shown twice.
Fits to the light curves overlay the histogram for phases 0 to 1 (blue in the electronic version), and the phase-aligned $\sim 1.4$ GHz radio profiles are also shown
(red curves). 
}
\label{WLCs_young}
\end{figure*}

\begin{figure*}[h]
\centering
\includegraphics[width=0.33\textwidth]{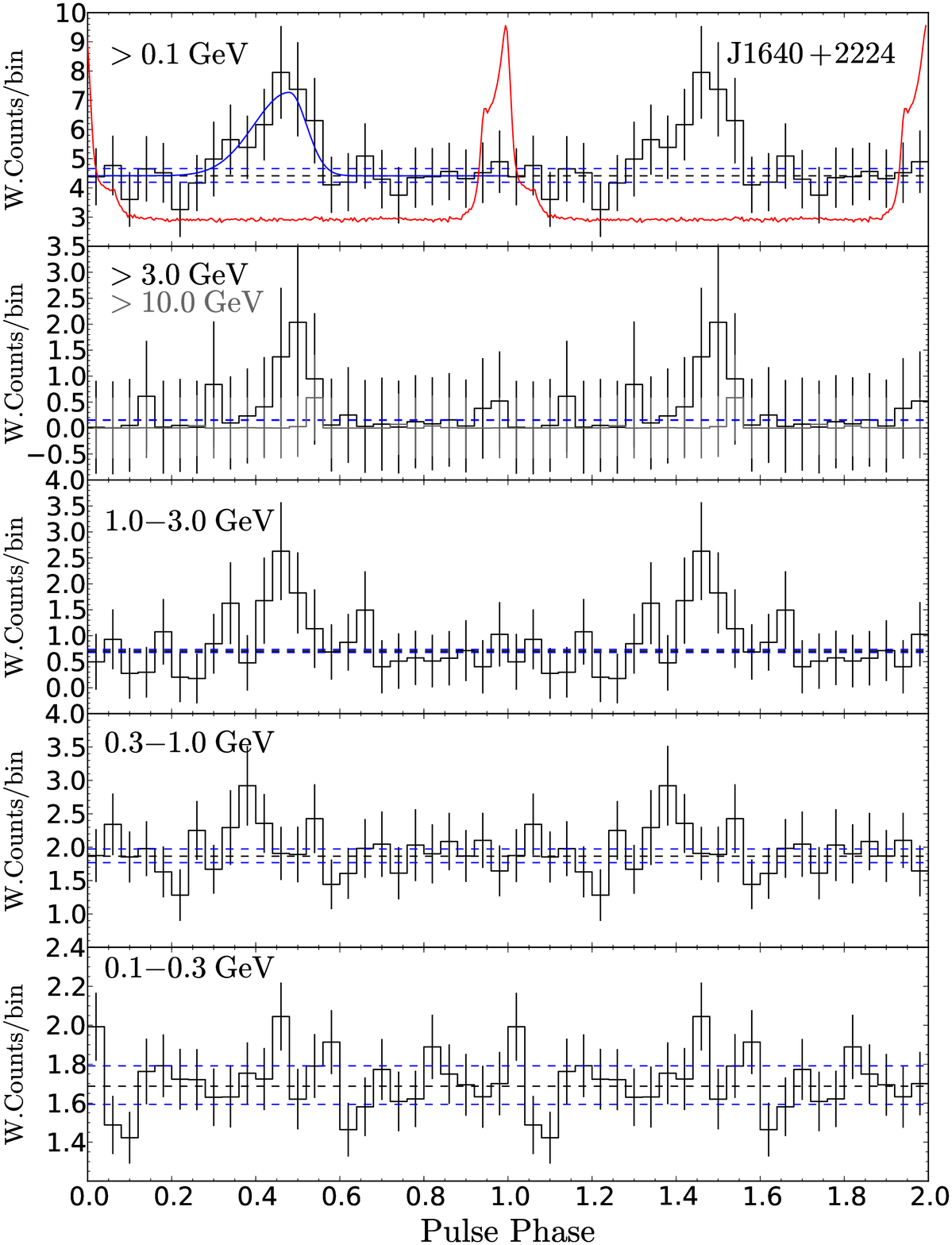}
\includegraphics[width=0.33\textwidth]{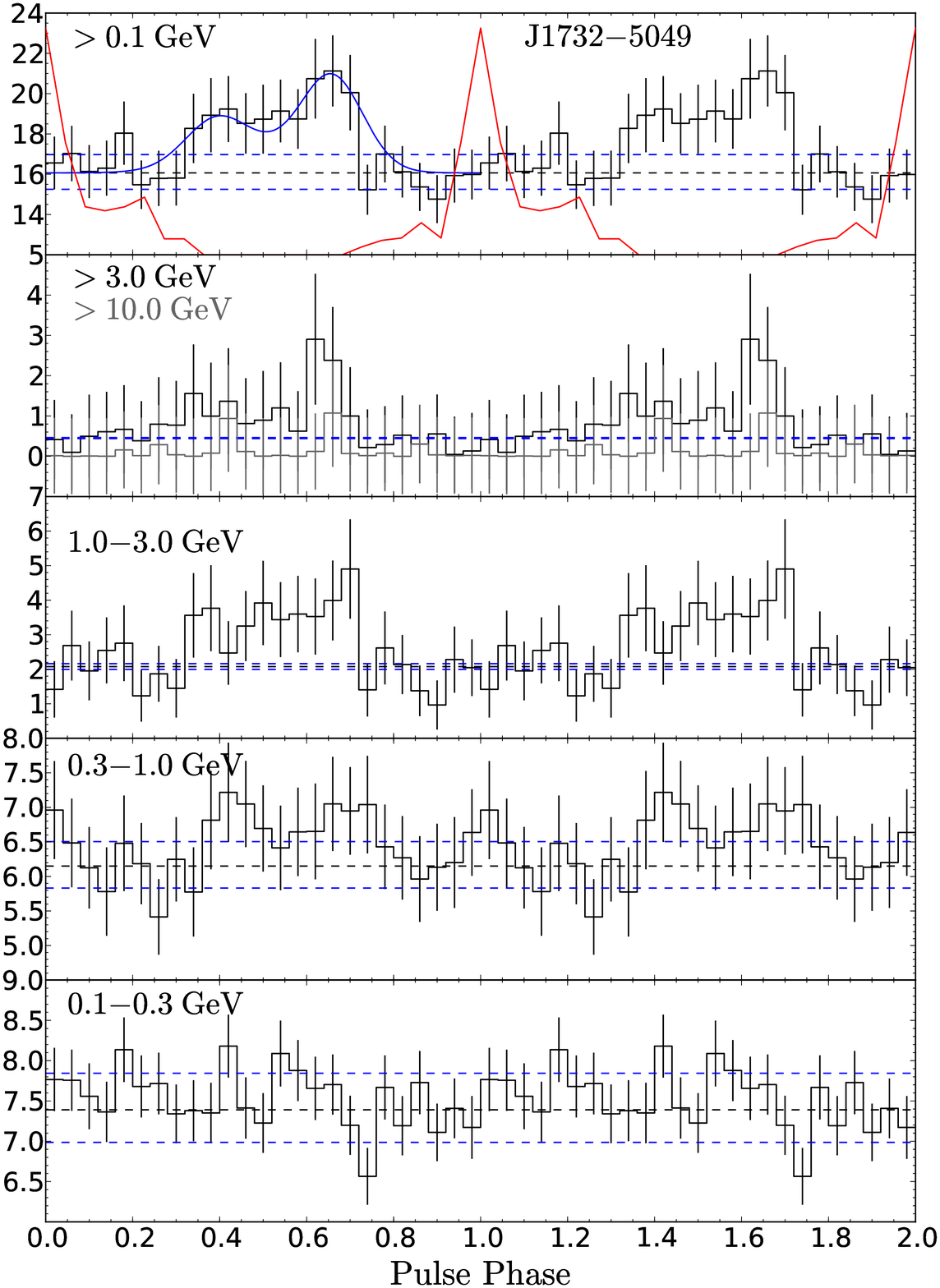}
\includegraphics[width=0.33\textwidth]{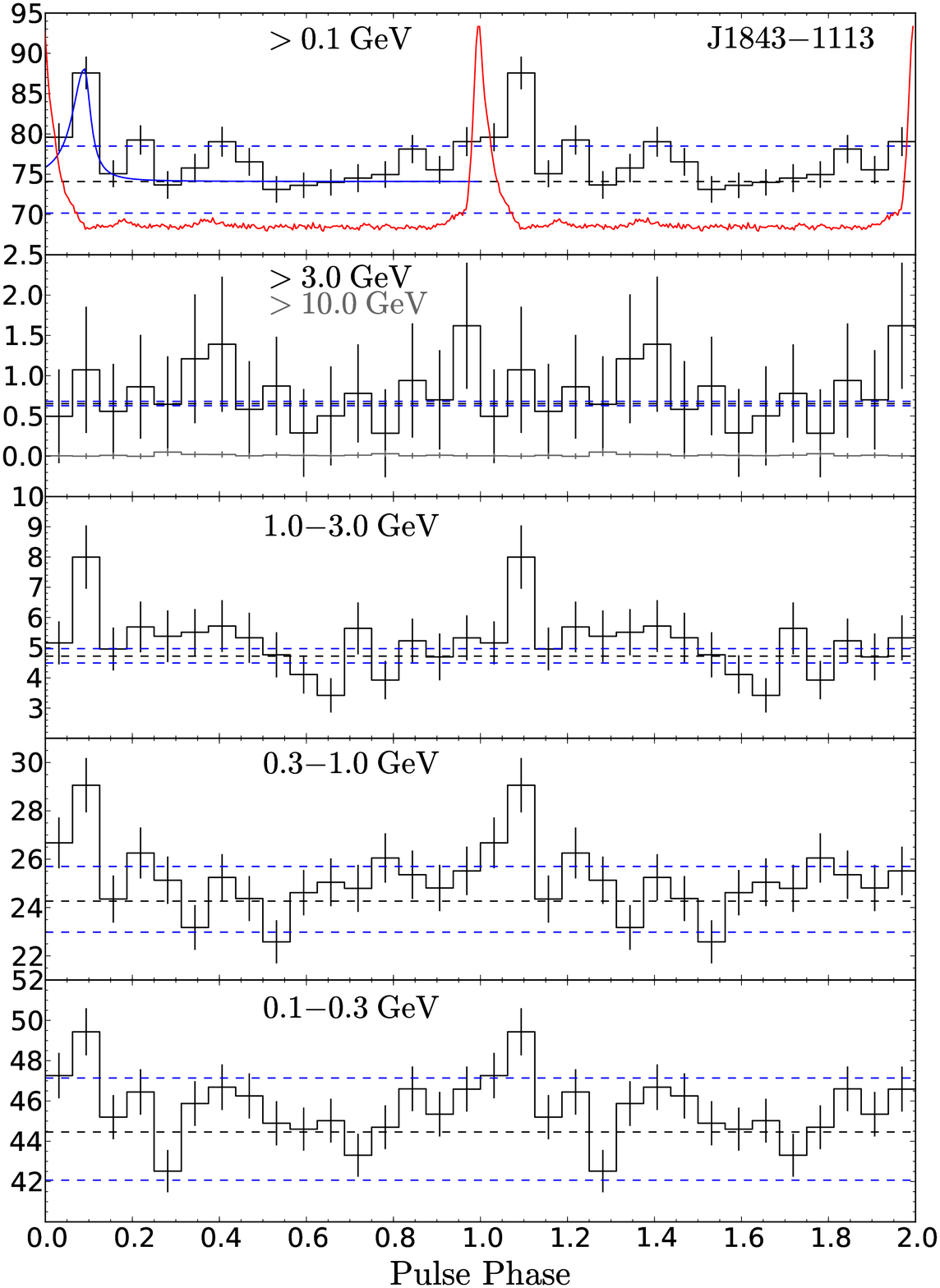}
\caption{Same as Fig. \ref{WLCs_young},  for millisecond pulsars PSR J1640+2224 (left), J1732$-$5049 (middle), and J1843$-$1113 (right). }
\label{WLCs_MSPs}
\end{figure*}

Gamma-ray photon phases were calculated with the \textit{fermi} plugin to the TEMPO2 pulsar timing software \citep{Edwards06}. 
The light curves in Figures \ref{WLCs_young} and \ref{WLCs_MSPs} use events from within $2^{\circ}$ of the pulsar radio position and are {\it weighted},
as described in Section $5.1$ of 2PC, except that the offset of $1\sigma$ reported there was not in fact used, neither in 2PC nor in this work.
The weights represent the probability that a photon comes from the pulsar rather than from nearby sources or the diffuse background, computed using the spectral results of Section \ref{specana}. 
The weighted $H-$test \citep{KerrWeighted} has improved sensitivity compared to the unweighted version \citep{deJager2010}. 
In the gamma-ray pulsation search we required $>5 \sigma$ significance to declare a detection.

We define zero phase at the pulsar's radio peak. We show 25 bins per rotation for the gamma-ray light curves, 
except for J1055$-$6028 (20 bins) and J1843$-$1113 (16 bins). 
We followed the fitting procedure presented in 2PC to characterize the observed gamma-ray profiles, exploring different
shape functions, and quantified by the maximum likelihood method. 
Table \ref{BigTab} lists some pulse profile parameters.

The offset $\delta$ between the radio and gamma-ray peaks depends on the extrapolation to infinite frequency of the radio pulse times-of-arrival using the DM value.
We determined the DM values as we built the rotation ephemerides used to phase-fold the gamma rays.They agree with
the published DMs listed in the ATNF pulsar database \citep{ATNFcatalog}\footnote{http://www.atnf.csiro.au/research/pulsar/psrcat/expert.html}.
The uncertainties on $\delta$ due to the DM uncertainties or to DM changes over time are negligible for the three MSPs.  
For the three young pulsars, our rotation model was ``whitened'' using WAVEs in TEMPO2, again with DM fixed to the published values.
For J1705$-$1906 the measured DM uncertainty and published rate of change \citep{Jodrell} are too small to affect our $\delta$ value.
For J1055$-$6028 and J1913+0904 the DM uncertainties translate to $\delta$ uncertainties smaller than our peak position uncertainties.
We did not explore in detail whether the DM changes over the data epoch for these two pulsars.

\begin{table*}[h]
\begin{center}
\caption{\small \textit{Fermi} LAT dataset. See Section 2 for definitions of terms.}
\begin{tabular}{ll}
\hline\hline
Time interval  &2008 Aug 4 to 2012 Dec 12 (MJD 54682.5 to 56273.5)  \\
Dataset   &{\it Reprocessed Pass 7}, ``Source'' event class. \\
IRFs      & P7REP\_SOURCE\_V15 \\
Energy band   & 100 MeV - 100 GeV   \\
Zenith angle cut     &$<100^{\circ}$\\
Rocking angle cut &$<52^{\circ}$ \\
ROI radius for spectral analysis     & $10^{\circ}$, off-center, except for PSR J1640+2224. \\
ROI radius for light curves     & $2^{\circ}$ \\
Galactic diffuse model   &{\it gll\_iem\_v05.fit} \\
Isotropic model    &{\it iso\_source\_v05.txt} \\
ScienceTools version    & v9r32p04 \\
\hline
\label{DataTab}
\end{tabular}
\end{center}
\end{table*}

\subsection{Spectra}
\label{specana}
\begin{figure*}
\centering
\includegraphics[width=0.9\textwidth]{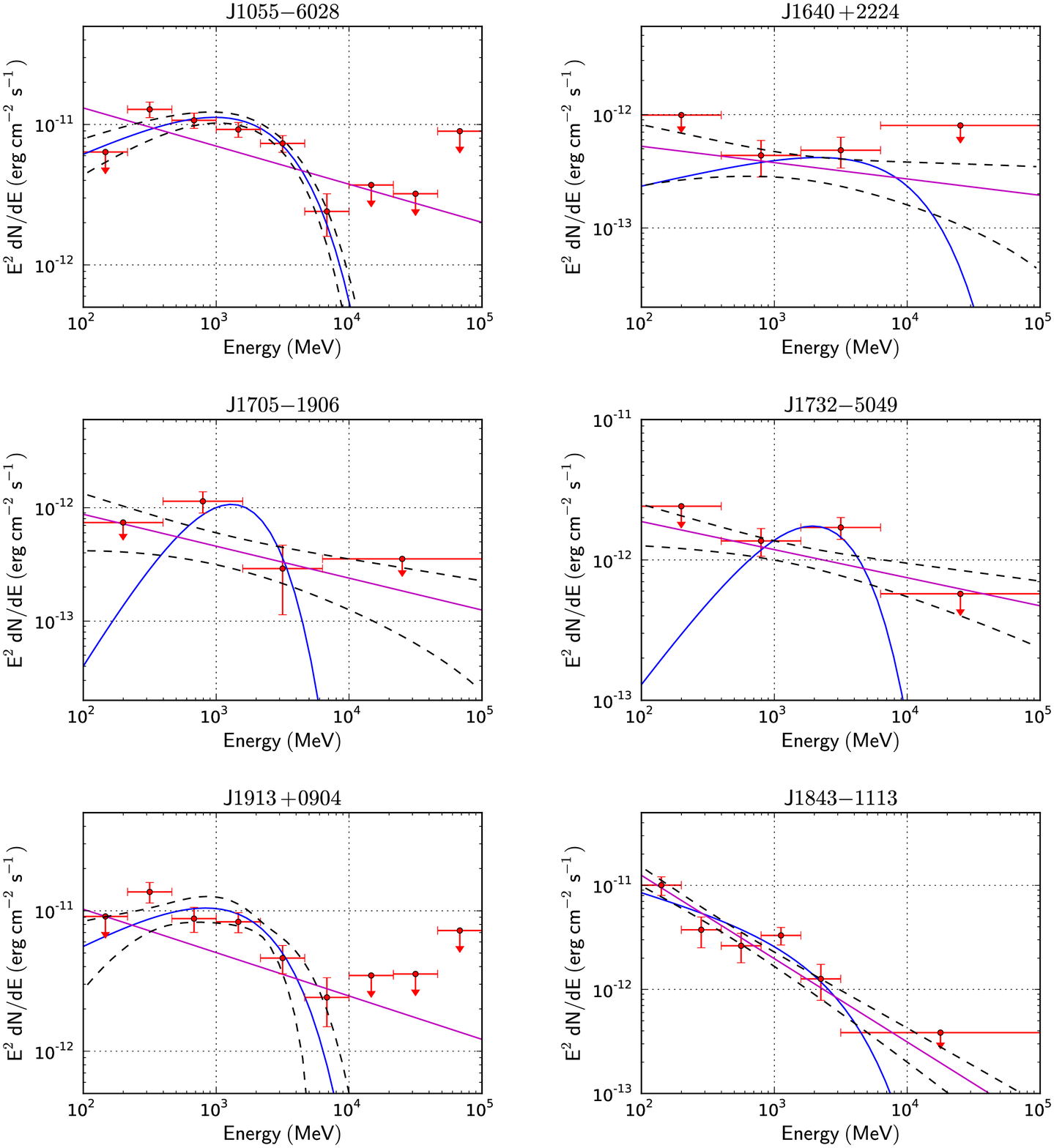}
\caption{Spectral energy distributions for the three young pulsars (left) and the three millisecond pulsars (right).
The solid straight lines are the power-law fits, whereas the solid curves show the exponentially cut off power laws. 
The dashed curves show the 95\% confidence level uncertainties for the fits reported in Table \ref{BigTab}. 
The data points are described in the text.
}
\label{SEDs}
\end{figure*}

Figure \ref{SEDs} shows the on-pulse spectral energy distributions (SEDs) of the six pulsars, and Table \ref{BigTab} lists the fit results. 
The pulsars are too weak for phase-resolved studies beyond the on-pulse selection described below.
We used the \textit{Fermi} Science Tool ``gtlike'', a maximum likelihood analysis tool that weights events from the target and background sources according to the energy-dependent PSF.
We used the binned analysis with the ``MINUIT'' optimizer.
The gtlike tool models data in a square region, which we set to $14^\circ$ on a side, inscribed within the offset $10^\circ$ radius data sample.

Instead of centering the ROIs at the pulsar positions, we shifted the ROI centers away from the bright sources.
The shifts allow us to obtain the smooth ``residuals maps'' that indicate that the modeling is reliable.
A residuals map shows the difference between spatial maps of observed gamma-ray counts, and the map of counts predicted by the likelihood model. 
The histogram of the residual map pixels has a gaussian distribution for well-modeled data yielding reliable spectral parameters. 
Mis-modeled sources cause zones of excess or deficit in the maps, and tails in the residuals histogram.
The pulsar's spectral parameters may or may not deviate from their best values in such cases.

The residuals maps for most of our pulsars had such tails, due to Galactic diffuse emission and bright nearby sources.
We therefore shifted the ROI centers to the positions listed in Table \ref{BigTab}.
As examples, we shifted PSR J1055$-$6028 away from the tangent of the Carina-Sagittarius spiral arm, where the diffuse emission is particularly bright.
Gamma-bright $\eta$ Car is $2^\circ$ away \citep{etaCarinae}.
We shifted PSR J1913+0904 away from the gamma-bright supernova remnant W44. 
PSR J1640+2224 did not need a shift.

The ``source models'' used by gtlike contain sources with ``Test Statistic'' $\rm TS\geq 16$ \citep{Mattox96} 
within a $15^\circ$ radius of the offset ROI center, taken from an interim 4-year source list.
Sources farther than 5$^\circ$ from the pulsar radio position, or outside of the offset $14^\circ$ square data region, 
were assigned fixed spectral forms with parameters taken from the 4-year list. 
Spectral parameters for the pulsars and remaining nearby sources were left free to refit. 
The Galactic diffuse emission model matched to our analysis version and IRFs is given in Table \ref{DataTab}, 
as is the isotropic component, which models both the extragalactic diffuse gamma-ray emission and residual misclassified cosmic rays. 
The source model .xml files will be available online at the FSSC$^1$.

PSR J1834$-$1113 has the largest ROI shift, nearly $5^\circ$. 
We verified that its spectral parameters with or without shifting the ROI are the same, within the systematic uncertainties described below.
PSR J1835$-$1106 is reported in 2PC but does not have $\rm TS\geq 16$ in the 4-year source list.
It is very close to gamma-bright PSR J1833$-$1034. 
Adding it to the source model does not change the spectral results.

 
The pulsars are modeled with an exponentially cutoff power law,
\begin{equation}\label{eq:spectra}
\frac{dN}{dE} = N_0 (E/E_0)^{-\Gamma} {\rm exp} \left[-\left(\frac{E}{E_{c}}\right)^{b}\right], 	
\end{equation}
where $N_0$ is the differential flux normalization (ph\,cm$^{-2}$\,s$^{-1}$\,MeV$^{-1}$), $\Gamma$ is the photon index, $E_{0}$ defines the energy scale, and $E_{c}$ is the cutoff energy. 
The 117 gamma-ray pulsars in 2PC are generally well-described by a simple exponential cutoff, $b=1$, a shape predicted by outer magnetosphere emission models.
The brighter pulsars show ``harder'' spectra, $b \approx 0.5$.
Pulsar brightness is not a concern in the current paper, and due mainly to the sources' weakness, freeing the $b$ parameter does not improve the fits.
We set $b = 1$.
We increase the signal-to-background ratio by selecting only photons in the on-peak intervals, defined in Table \ref{BigTab}. 
The fit results are the same within statistical uncertainties whether we select by phase or not, but the cut improves the TS, that is, the significance. 

The spectral cutoffs are robust for only two of the young pulsars, and Table \ref{BigTab} lists their cutoff energies $E_c$.
For the other four pulsars the difference of the log likelihood fits with $b=0$ versus 1 is TS$_{\rm cut}<13$. 
That is, we cannot reliably distinguish a pure power law from one with a cutoff, and tabulate only the simpler spectral shape.
The energy flux integrated above 100 MeV, $G_{100}$, is more robust than the integrated photon flux, 
because the detector acceptance increases and the PSF narrows at higher energies \citep{P7Paper}.
We report only $G_{100}$.

Figure \ref{SEDs} shows both the power-law ($b=0$) and cutoff ($b=1$) fits, showing uncertainties (dashed curves) only for the functional form reported in Table \ref{BigTab}.
The points in the figure come from a maximum likelihood analysis in logarithmically-spaced energy bands between $0.1$ and 100 GeV.  
Each pulsar is modeled by a simple power law with index $\Gamma=2$ in each band, with all other sources fixed to their nominal parameters,
except that the normalization of the Galactic diffuse component is left free for the three low latitude sources.
Upper limits (95\% confidence level, calculated using the Bayesian method) on the pulsar flux are shown if the pulsar has $\rm TS<4$ in a given band. 

Systematic uncertainties were estimated by re-fitting the on-peak data with the Galactic diffuse level shifted by its approximate uncertainty of $\pm 6\%$, 
and by re-fitting the on-peak data using effective area functions $A_{eff}$ that ``bracket'' the uncertainty range \citep{OnOrbit}.
The overall systematic uncertainties that we quote are the sums in quadrature of the parameter shifts resulting from changing 
the effective area and the Galactic diffuse intensity.
The discussion of 2PC Equation 14 explains the method.
The smooth interpolation of $A_{eff}$ versus $\log (E)$ yields $\pm$ 10\% at 0.1 GeV, $\pm$ 5\% near 0.56 GeV, and $\pm$ 5\% at 10 GeV. 

An improvement made here as compared to 2PC concerns the treatment of the sources more than $5^\circ$ from the pulsar, for which the spectral parameters are fixed. 
The numbers of predicted counts from each of the outlying sources were obtained using the nominal $A_{eff}$.
The likelihood fit with the modified $A_{eff}$  uses these predicted counts to determine the parameters of the nearby sources, and of the pulsar.
This ensures that the total number of predicted counts matches the number observed in the data.
Table \ref{BigTab} shows large systematic shifts of the spectral parameters in a few cases, not surprising for signals just emerging above the background,
but overall the statistical and systematic uncertainties have the same magnitude and we find that the results are robust. 

\subsection{Distances}

Table \ref{BigTab} lists the pulsar distances. 
None of the pulsars in this paper are in the ongoing campaign to measure pulsar parallaxes with the
Very Long Baseline Array \citep[VLBA, see][]{cbv+09,Chatterjee2013}. 
We discuss the timing parallax distance of PSR J1640+2224 at the end of this section.
The other five distances were obtained using the DM 
and the NE2001 model of the electron density throughout the Milky Way \citep{Cordes2002}.
As in 2PC, we estimate the distance uncertainties by re-running NE2001 for the two values $(1\pm0.2)$DM, except for PSR J1055$-$6028, discussed below.

\begin{figure*}
\centering
\includegraphics[width=0.49\textwidth]{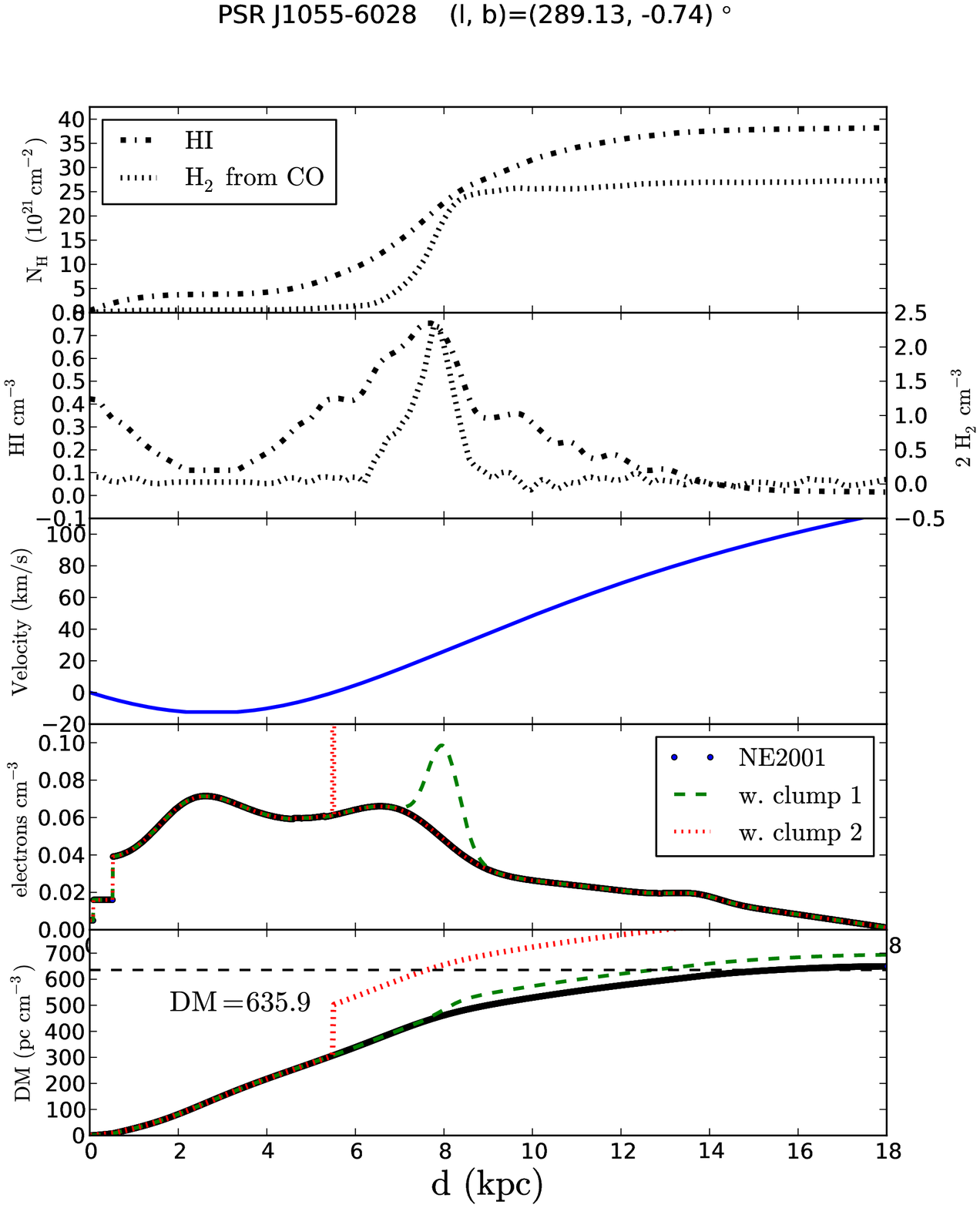}
\includegraphics[width=0.49\textwidth]{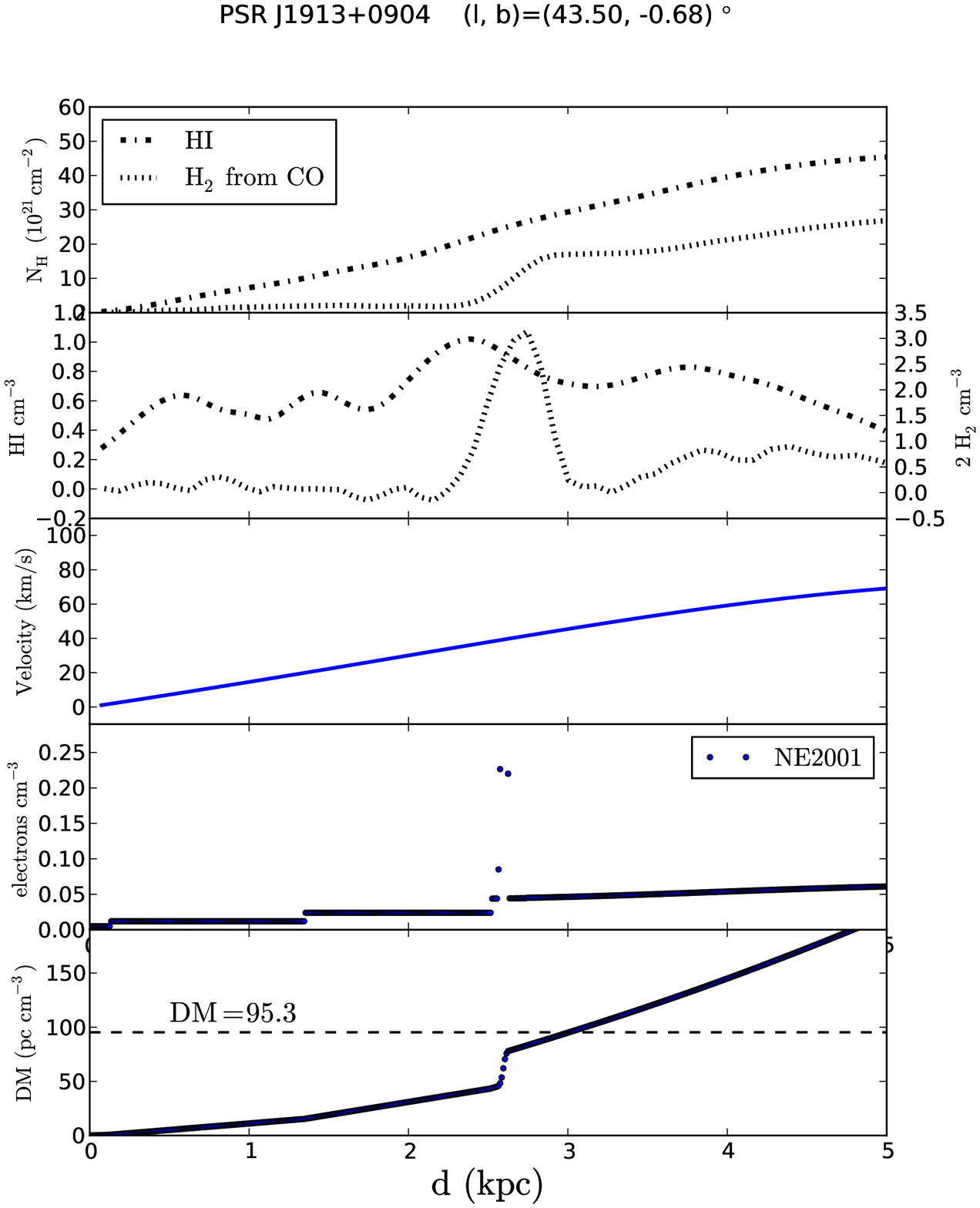}
\caption{Diagnostics for DM distances obtained using the NE2001 model, for PSRs J1055$-$6028 (left) and J1913+0904 (right).
Counting from the top, the second frame shows the proton densities along the line of sight of the pulsar derived from the LAB survey HI brightness temperature $T_b$
(left axis), and from the \citet{DameCOsurvey} CO survey brightness temperature (right axis). 
The radial velocity from the surveys has been translated to a distance (x-axis for all frames) using a flat rotation curve (third frame).
The top frame shows the proton densities integrated into column densities.
The fourth frame is the electron density used by NE2001, integrated to give DM in the fifth frame (the spikes exceed the y-scale).
See text for a discussion of the clumps added to the NE2001 model for J1055$-$6028.
}
\label{ColDensities}
\end{figure*}

NE2001 is the principal tool used to obtain radio pulsar distances, providing useful, even accurate, distances for most radio-loud pulsars.
``Accurate'' is cited as $\pm30$\% by some authors, although evidence for that number is uneven.
\citet{SchnitzelerElectrons} finds that modeled DMs match observed DMs to within a factor of 1.5 to 2 for about 75\% of the directions to pulsars.
Examples of the NE2001 distance being wrong by a factor of several include:
PSR J0248+6021, behind a dense part of the Heart and Soul nebula \citep{Theureau2011};
PSR J0855$-$4644, with a line of sight tangent to the Gum nebula \citep{Acero2013};
and PSR J2021+3651 \citep{PSR2021LAT}, in the Cygnus region.
Uncertainty about which distances are or are not reasonable casts doubt on gamma-ray pulsar luminosity and space distributions in general. 
To cross-check the DM distances, useful even with the forthcoming improvements to NE2001 \citep{NEv3p0},
we compare the NE2001 electron density $n_e$ along the lines of sight to the pulsars with different observations.
Figure \ref{ColDensities} illustrates these ``diagnostics'' for the two pulsars with the largest distances in our sample.

The first is the atomic hydrogen (HI) number density. It is, on average, ten times higher than the electron number density \citep{HeNgKaspiDMvsNH}.
The ratio is lower when, for example, nearby OB stars provide intense ionizing ultraviolet light. 
The HI number density is $1.823 \times 10^{18} T_b \Delta V/\Delta d$ where the constant is obtained assuming that the HI is optically thin at 21 cm \citep{GalacticHI},
and $T_b$ is the brightness temperature recorded by the Leiden-Argentine-Bonn (LAB) survey \citep{Kalberla2005}.
The survey provides $T_b$ in steps of $v_{LSR}$, where $v_{LSR}$ is the Doppler shifted line-of-sight velocity with respect to the local standard of rest. 
For HI in the Milky Way, a large component of $v_{LSR}$ is due to the rotation of the Galaxy. 
We approximate the rotation curve as flat, with a rotational velocity of 235 km s$^{-1}$ and a distance of 8.5 kpc between
the Sun and the Galactic center. 
We derive HI densities under the assumption that the spectral shifts are due entirely to Galactic rotation and that for the channel width of the
survey $\Delta V$, the gas is uniformly distributed along the line of sight over a distance $\Delta d$ converted from $\Delta V$ using the rotation curve.

Carbon monoxide (CO) traces molecular hydrogen, H$_2$. The fraction of electrons in a molecular cloud is very small, of order $10^{-7}$ \citep{ElectronsInClouds}.
Nevertheless, CO indicates the presence of dense, molecular gas and, perhaps, of an electron excess. 
We use the CO survey by \citet{DameCOsurvey} and a similar conversion for $T_b$ to hydrogen molecule density,
$2\times 1.8 \times 10^{20} T_b \Delta V/\Delta d$.  
Figure \ref{ColDensities} indicates clouds on the lines of sight for both pulsars.
For PSR J1055$-$6028 the two possible distances for a given $v_{LSR}$ are shown: 
when a distance corresponds to two $v_{LSR}$ values, the quantity derived from the HI and CO radio intensity is plotted twice, once for $d<3$ kpc and again for $3 < d < 5.5$ kpc.
At larger distances in this direction there is no ambiguity.
For PSR J1913+0904, the mirror point is at a distance incompatible with the DM and is not shown.
At high latitudes ($|b| > 5^\circ$) the CO survey data is integrated over $v_{LSR}$ and the distance conversion is not available.

Finally, we also examine the Doppler-shifted H$_\alpha$ hydrogen recombination line intensity, 
which also indicates the presence of electrons, as a function of $v_{LSR}$, but did not include it in Figure \ref{ColDensities}.
We use H$_\alpha$ spectra from the Wisconsin H-Alpha Mapper (WHAM) survey \citep{WHAM}, 
and the Southern H-Alpha Sky Survey Atlas (SHASSA) for southern sources \citep{SHASSA}.
Both surveys, as well as the compilation by \citet{FinkbeinerHalpha}, provide images integrated over $v_{LSR}$ showing
structures that may be on or near each pulsar's line of sight. The image intensities correspond to emission measures
EM $\approx 2.25\, R$, with EM in units of pc cm$^{-6}$ and $R$ in Rayleighs, assuming a gas temperature of 8000 K. 
The NE2001 model estimates EM from the integral of $n_e^2$ along the line of sight, which is sensitive to the non-uniformity of $n_e$.
We do not compare the measured and predicted EM values in this work, but do note the detailed study of different Galactic electron models using the H$_\alpha$ maps
by \citet{SchnitzelerElectrons}.

NE2001 models electron densities in e.g. the arms and halo of the Galaxy as being uniform, 
with correction tables of ``clumps'' and ``voids'' for directions where anomalies have been noticed.
We search for unmodeled anomalies. 
The difficulty is knowing what size and density to assign to a new clump or void.
Exchanging the ``standard'' NE2001 biases with our own {\it ad hoc} variants would add only confusion for small distance shifts. 
\textit{In fine}, we added an additional clump along the line of sight only for PSR J1055$-$6028.

PSR J1055$-$6028 was discovered in reprocessing of the Parkes multibeam pulsar survey (PMPS) data.
Improved algorithms picked out signals missed during earlier analyses \citep{J1055-6028_discovery}, 
adding to the total of over 750 pulsars discovered in the survey.
At the nominal NE2001 distance of $15.5^{+3.5}_{-6.3}$ kpc, it would match the farthest detected gamma-ray pulsar (PSR J1410$-$6132 at
an NE2001 distance of 15.6 kpc).
Figure \ref{ColDensities} (bottom frame, left) shows this, where the integrated electron curve reaches the measured DM value of 636 pc cm$^{-3}$.
Figure \ref{ColDensities} (left) also shows an excess of both HI and CO around 8 kpc in that direction.
Both \citet{AnatomicallyCorrect} and NE2001 show the line of sight running through (tangent to) the Carina-Sagittarius spiral arm in that range,
entering (exiting) the arm at about 7 (9) kpc.
Two massive stars appear near the pulsar (see also Section 3.2). 
AG Carinae (AG Car) is a 50 solar-mass luminous blue variable, $0.07^\circ$ degrees away, with a ring nebula of angular size $0.01^\circ$.
\citet{AGcarDist} place AG Car at $6 \pm 1$ kpc from Earth.
GG Carinae (GG Car) is a $\beta$ Lyrae-type eclipsing binary, $0.09^\circ$ degrees away, at $5 \pm 1$ kpc \citep{GGcarDist}.
Both angular separations correspond to 7 or 8 pc from the line of sight.

The SHASSA H$_\alpha$ image shows high EM (400 pc cm$^{-6}$) in the direction of the pulsar, consistent with clumps of electrons along the line of sight. 
AG Car is five times brighter and GG Car is three times brighter, illuminating the region (in projection) between the stars and extending to the pulsar.
We thus propose two ``clump'' scenarios. The first appears in the Figure as the $n_e$ bump at 8 kpc.
It raises $n_e$ to $0.1$ electrons cm$^{-3}$, an eighth of the HI density at the distance of the HI and CO peak intensities.
The size of $0.5$ kpc is based on the extent of the HI cloud, amongst the largest of NE2001 clumps.
This only decreases the DM distance to 13 kpc, within the uncertainty obtained from the $(1\pm0.2)$DM prescription.

In the second scenario, we place the electron clump between the stars (\textit{l,b} = $289.17, -0.68$ and 5.5 kpc), with a radius of 10 pc (typical in NE2001).
We ``reverse engineer'' the density. That is, we find that the DM yields 6.8 (8.7) kpc for $n_e =$ 25 (15) electrons cm$^{-3}$, respectively.
The Figure shows 7.6 kpc for $n_e =$ 20 electrons cm$^{-3}$. 
In NE2001, $n_e$ decreases exponentially with angular distance from the center of the clump, truncated at the clump radius.
The spike in the Figure reaches 11 electrons cm$^{-3}$, off-scale for readibility.
The largest NE2001 clump densities are 40 electrons cm$^{-3}$. The densities necessary for such a large DM at half the standard distance are realistic.
We express a prejudice in favor of PSR J1055$-$6028 being in the tangent region of the arm, at the distance of the massive stars, but allow for it to
be at the distance suggested by HI alone, and will use $8^{+5}_{-3}$ kpc. It remains one of the more distant gamma-ray pulsars.

PSR J1913+0904's nominal NE2001 distance of $3.0 \pm 0.4$ kpc is typical of the gamma-ray pulsars reported in 2PC.
The H$_\alpha$ intensity at its position is near the background level (a few pc cm$^{-6}$).
Figure \ref{ColDensities} shows a dense CO cloud near 2.6 kpc. 
The $n_e$ spike at the same distance comes from the line of sight grazing a standard NE2001 clump modeling the methanol maser CH3OH 43.80-0.13,
with $n_e = 6$ electrons cm$^{-3}$.
The maximum density due to the maser clump, $0.85$  electrons cm$^{-3}$, is off the plot scale.
Away from the spike the density is  $n_e < 0.05$ electrons cm$^{-3}$, while the HI density is $\sim 0.9$ atoms cm$^{-3}$.
Since on average, as stated above, HI/$n_e \approx 10$, $n_e$ away from the spike could be larger. 
The pulsar would then be nearer, but in fact the distance prediction is especially sensitive to the details of the maser modeling,
and we retain the NE2001 value.
The other four pulsars are nearer than 2 kpc, with no striking features in the different maps. 
In general, the maps better indicate possible $n_e$ mismodeling for more distant pulsars, beyond the large structures near the solar region,
and where radial velocities are higher.

The timing parallax for PSR J1640+2224 corresponds to $450^{+660}_{-160}$ pc, less than half the NE2001 value, although the uncertainties overlap.
The Lutz-Kelker correction provided by \citet{Verbiest12} accounts for the bias skewing observed distances away from the true values.
For PSR J1640+2224 it yields a very close distance, $<400 $ pc, 
a consequence of the large parallax uncertainty, and we do not use the correction. 
The NE2001 $n_e$ versus distance curve shows a discontinuity in the first kpc, similar to that shown for PSR J1055$-$6028,
followed by a steep decline in density as the line of sight for this high-latitude pulsar leaves the Galactic disk.
Smoothing it, and doubling the pre-step density as suggested by the HI curve, would surely bring the DM distance into agreement
with the parallax measurement.

X-ray observations of absorption can yield hydrogen column densities, which can be compared with the integrated proton densities
from the HI, CO, and H$_\alpha$  measurements for an independent distance estimate.
We searched the archives and found 28 ks and 8 ks observations with XMM-Newton 
for PSRs J1055$-$6028 (PI: Y.~ Naz{\'e}) and J1705$-$1906 (PI: K.~Mason), respectively,  
and 9 ks with Swift (from December 2011 - December 2013) for J1640+2224 \citep{SwiftJ1640}.
The observations cover $0.3$ to several keV. 
The PSR J1055$-$6028 data were taken in the field of an observation targeting AG Car which resulted in a non-detection of the star \citep{AGcarXray}.
Analysis reveals no X-ray counterparts for any of the three pulsars, and thus, no distance constraints. 
A 20 ks Chandra observation of J1843$-$1113 (PI: K. Wood) used the High Resolution Camera and is unsuitable for such spectral analysis.



\subsection{Gamma-ray luminosity}
The luminosity is $L_\gamma = 4 \pi f_\Omega G_{100} d^2$.
The ``beaming factor'' $f_\Omega$ is the ratio of the power radiated by the pulsar into all space to the power radiated towards a given line of sight
(see Eq. 16 in 2PC). 
Restating, $f_\Omega$ normalizes the observed intensity for a given inclination $\zeta$ of the pulsar rotation axis to the average intensity over all inclinations. 
Emission models predict $f_\Omega \approx 1$ for our sample, and we set $f_\Omega=1$.


Table \ref{BigTab} includes the luminosities and the efficiencies $\eta = L_\gamma / \dot E$ obtained for each pulsar.
The systematic uncertainties include two contributions: the flux uncertainty due to uncertainties in the effective area and the diffuse emission ;
and the propagated distance uncertainties. We added these uncertainties in quadrature.

We adopted the widely-used value for the neutron star moment of inertia, $I = 10^{45}$ g cm$^2$, to calculate $\dot E = 4\pi^2 I \dot P P^{-3}$.
Much evidence indicates that neutron star radii and masses are larger than the 10 km and $1.4$ M$_\odot$ used to obtain this value of $I$ \citep{NS_EoS}.
The moments of inertia are probably twice as high for young pulsars, and perhaps three times larger for recycled pulsars.
The efficiencies would then be smaller by the same factors.

%
%
%
%

\section{Six Faint Gamma-ray Pulsars}
\label{indipsrs}
\subsection{Identifying Selection Bias}
``Faint'' means more than just having a low flux. 
Several parameters affect how many years of LAT observations it takes to detect a given pulsar.
Selection bias thus affects the \textit{Fermi} pulsar sample at some level, in spite of the uniformity of the all-sky survey.
Identifying these biases is a first step. In the next section, we begin calculations to correct for the bias due to
profile shape, with the aim to allow more accurate comparisons with population syntheses.

A first example of parameters affecting detection is provided by
PSR J1843$-$1113, discovered in the PMPS \citep{Hobbs_pmbs}.  
Its distance (1.7 kpc) is large but not unusual for gamma-ray MSPs.
Its $\dot E$ is high for an MSP, while its latitude is low for an MSP but mid-range for young pulsars.
In fact it is one of the gamma-ray MSPs closest to the plane, giving it a high diffuse emission background level. 
(Figure 17 of 2PC shows the latitude dependence of the LAT detection sensitivity for unpulsed sources with pulsar-like spectra.) 
Figure \ref{SEDs}, however, shows what is probably the dominant effect:
the spectral index $\Gamma = 2.8$ is the steepest of any gamma-ray MSP and, crucially, as steep as the diffuse emission spectrum. 
Thus, whereas most pulsars rise above the diffuse background in some spectral range, J1843$-$1113 does not.

PSR J1732$-$5049 was discovered in a Parkes survey of intermediate Galactic latitudes at 1.4 GHz \citep{midLatHiFreqSurvey}. 
It is slightly closer, and farther off the plane, than J1843$-$1113 but has twenty times smaller spindown power. 
Luminosity falls more steeply with decreasing $\dot E$ for low spindown powers.
This pulsar was not in the second LAT source catalog (2FGL), but
a source positionally consistent with the pulsar was found during the development of the 3FGL catalog based on four years of LAT data.
We obtained a Parkes rotation ephemeris\footnote{R.N. Manchester, personal communication.},
based on observations described by \citet{PPTA}.
Phase folding the gamma-ray data immediately yielded compelling pulsations: pulsed searches are generally more sensitive,
but an ephemeris is needed. 

Two more pulsars in our sample have very low $\dot E$: PSRs J1640+2224 and J1705$-$1906.
Our parallax distance gives J1640+2224 a luminosity as low as the two previous record holders, MSPs J0437$-$4715 and J1024$-$0719 (2PC).
PSR J1640+2224 was discovered in a high latitude search at Arecibo \citep{J1640discovery}.
Its low spindown power and luminosity are balanced by favorable distance and latitude.
PSR J1705$-$1906 was one of 155 pulsars discovered during the second Molonglo survey of the entire sky south of declination $20^\circ$ \citep{Molonglo},
and is in one of the ``\textit{Fermi bubbles}'' \citep{FermiBubbles}, so that the diffuse background level is higher than might otherwise be expected.
Its nearness helps make it detectable.
J1705$-$1906 has the lowest spindown power and lowest luminosity of any young, radio-loud gamma-ray pulsar to date. 
The spectral cutoff is at the lowest energy of our sample, difficult to measure due to the weak signal-to-noise ratio.

Close examination of the $P\dot P$ diagram (Fig. 1 of 2PC) reveals that just below the ``empirical death-line'' discussed in 2PC,
between $1$ and $3\times 10^{33}$ erg s$^{-1}$, very few pulsars were gamma-ray phase-folded. 
Work is currently in progress to obtain ephemerides, primarily from Jodrell Bank Observatory, to better explore this
niche in parameter space and in particular, to see whether the $3\times 10^{33}$ value reflects a pulsar property or selection bias.

A large beaming factor, $f_\Omega$, gives a small flux for a ``normal'' luminosity, as discussed by \citet{subluminous} for ``sub-luminous'' pulsars.
Gamma-ray emission models can \textit{predict} $f_\Omega$ for a given pulsar only if its inclination $\zeta$ as well as the angle $\alpha$ between the
rotation and magnetic axes are known. 
Without the prediction the \textit{a priori} detectibility of a given pulsar is in doubt, 
although in practice $f_\Omega$ rarely changes the flux by a large factor.  
Once a pulsar has been detected, $(\alpha, \zeta)$ and hence $f_\Omega$ can be estimated by comparing the observed profiles with model predictions. 
\citet{AtlasII} provided an ``Atlas'' of predicted profile shapes over a grid of inclination angles for young pulsars, for three emission models. 
For simplicity, we refer mainly to the ``outer gap'' or OG model, well-suited to PSR J1705$-$1906, discussed below.
The grid cells have as many as four profiles, corresponding to small to large widths $w \equiv \sqrt{\dot E}$ of the outer gap.
MSPs have much smaller light cylinders, changing the shapes of the gamma-ray emission zones compared to young pulsars. 
\citet{2PC_MSPmodeling} have modeled all of the 2PC gamma-ray MSPs.

Another way to estimate $(\alpha, \zeta)$ is to analyze the radio polarization position angle (PPA) variation with rotation phase \citep{CraigRomaniRVM}.
PSR J1705$-$1906 stands out in our sample as only the fourth young gamma-ray pulsar with a radio interpulse, 
visible in the top-middle frame of Fig. \ref{WLCs_young}. 
The previous are the EGRET pulsars Crab and PSR B1055$-$52, and 2PC pulsar J0908$-$4913.
In principle, sampling the PPA swing at two widely separated phase intervals should constrain $(\alpha, \zeta)$ better than for pulsars with a single narrow
radio peak. We thus expected PSR J1705$-$1906 to allow detailed gamma-ray modeling, and firm predictions for $f_\Omega$. 
However, a `kink' in the PPA swing of J1705$-$1906's interpulse is incompatible with simple models.
The kink is examined by \citet{B1702IP}, 
along with a wealth of other radio observations showing, for example, that the two radio emission regions are physically connected.
Of the three distinct scenarios they propose, none explain all of the data. 

For J1705$-$1906's very low $\dot E$, the observed single-peak gamma-ray profile is expected only for the Atlas OG model for $\alpha \gtrsim 50^\circ$. 
Pulsar radio beams will be seen from Earth only if $|\zeta - \alpha| \lesssim \rho$, 
where the half-angle of the radio cone is typically taken as $\rho = 5.8^\circ P^{-1/2}$ \citep{PopYoungGamma}.
For the period of J1705$-$1906 this gives $\rho \approx 10^\circ$. 
An independent analysis of the radio pulse widths by \citet{B1702IP} yielded $\rho = 12^\circ$, 
assuming that the pulsar is an orthogonal rotator.
We thus find $\zeta < 60^\circ$, and the Atlas then favors even larger $\alpha$ values.
That is, J1705$-$1906 appears confirmed as an orthogonal rotator, supporting the two-pole and the bidirectional scenarios discussed by \citet{B1702IP}.

\subsection{Profile Shapes and Detection Sensitivity}
Striking in Figures \ref{WLCs_young} and \ref{WLCs_MSPs} is the absence of any `classic' gamma-ray pulse profiles. 
None of the six resembles the Crab or Vela's lightcurves with two narrow peaks separated by a half rotation.
Even when Table \ref{BigTab} lists two peaks in the profile fit, 
they are so close together that they appear as a single, broad, uneven pulse, perhaps with a small peak off to the side. 

PSRs J1055$-$6028 and J1913+0904 are the most extreme cases.
PSR J1913+0904 was discovered with the PMPS a few years before the launch of \textit{Fermi} \citep{lfl+06}, and has since received little attention in the literature.
Both pulsars have large spindown powers and lie near the Galactic equator, in busy regions.
The two pulsars are the farthest of our sample, although J1913+0904 is `only' 3 kpc away.
Their spectral parameters are very typical.
Their ``\textit{faintness}'' comes from having larger pulse duty cycles than any 2PC pulsar: 
in Figure \ref{WLCs_young}, over 60\% of the $>100$ MeV pulse profile is above the background $+1\sigma$ level.
The largest duty cycles in 2PC are 40\% to 50\%. 

\begin{figure*}
\centering
\includegraphics[width=0.49\textwidth]{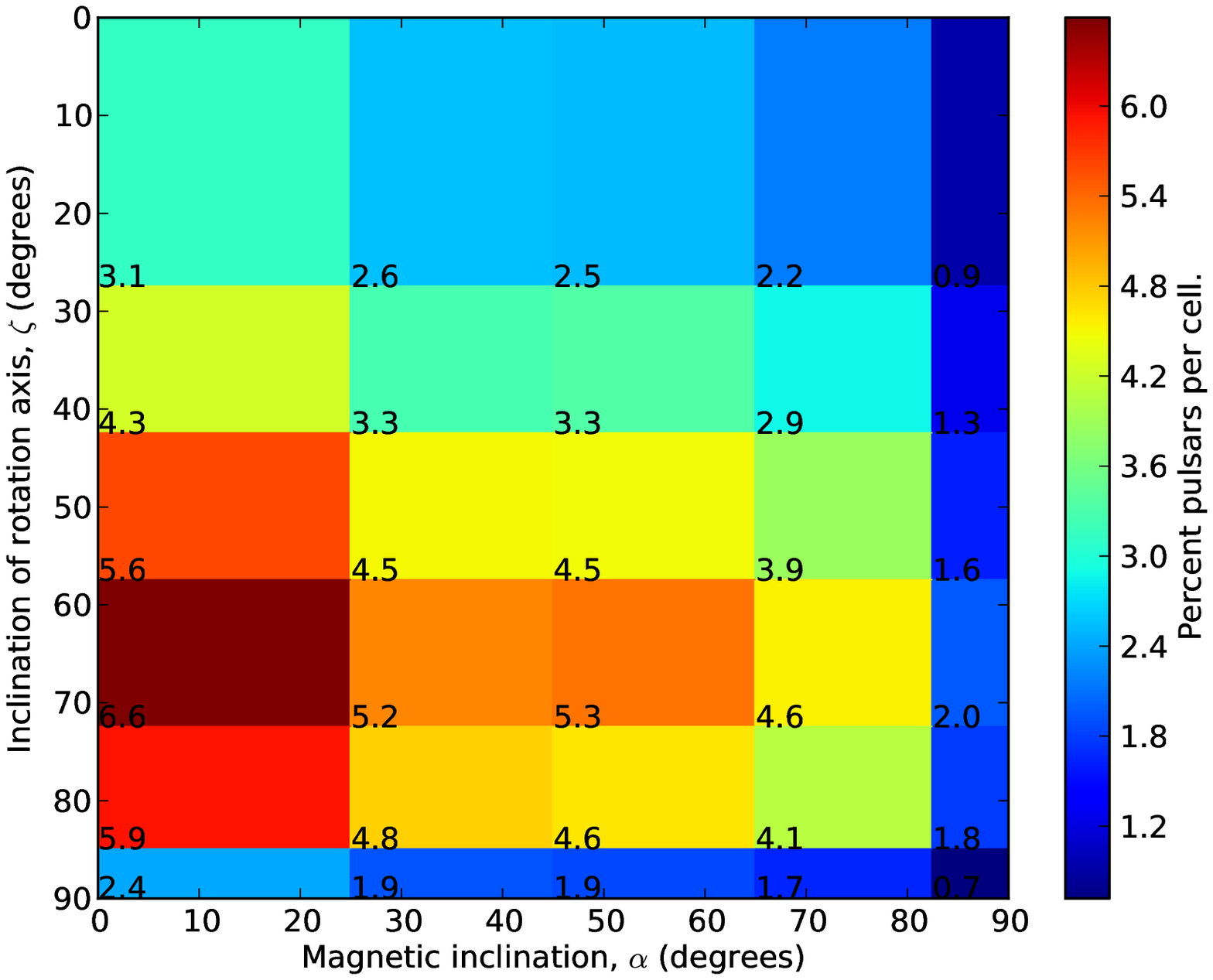}
\includegraphics[width=0.49\textwidth]{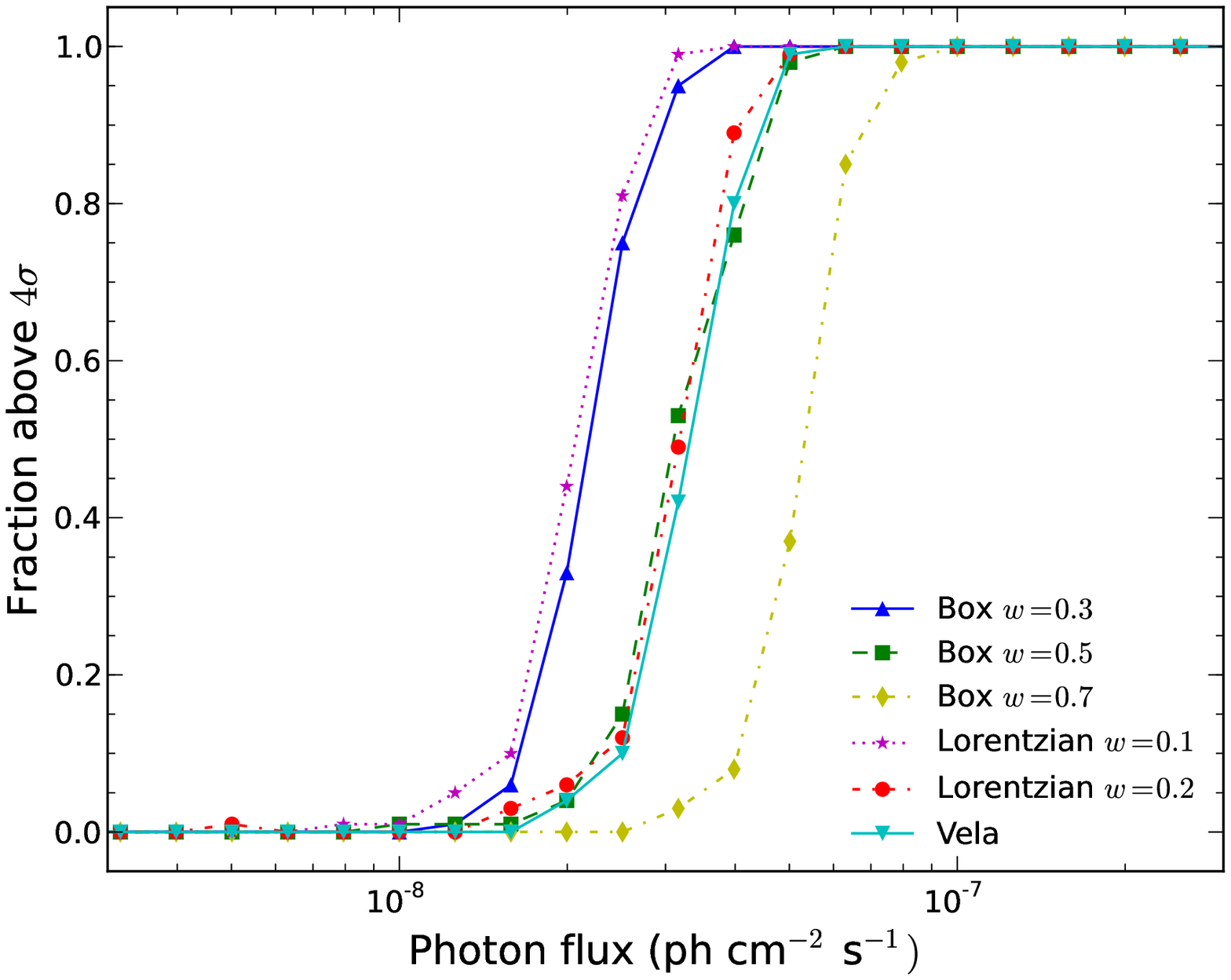}
\caption{Left: The same gridding in inclination $\zeta$ and magnetic inclination $\alpha$ as
in the Atlas of gamma-ray profile shapes of \citet[Fig. 15, ][]{AtlasII}. 
The occupancy fractions assume uniform distributions in $\alpha$ and $\cos\zeta$. 
Right: Fraction of simulated lightcurves exceeding $4\sigma$ with the H-test, as a function of pulsar flux above 100 MeV,
for a fixed background intensity, for different profile shapes. $w$ is the gap width defined in the text.}
\label{AlphaZeta}
\end{figure*}

To provide an indication of how common or rare broad-peaked pulsars might be,
Figure \ref{AlphaZeta} has the same $(\alpha, \zeta)$ grid as Fig. 15 of the Atlas.
Radio-loud pulsars favor the top-left to bottom-right diagonal of the grid 
(the magnetic axis passes near the line of sight, $(\zeta - \alpha) \lesssim \rho$, as above). 
`Classic' gamma-ray pulsars dominate the lower-right cells. The upper-left cells correspond to radio-loud, gamma-quiet pulsars. 
Broad peaks as for PSRs J1055$-$6028 and J1913+0904 appear at the upper-right: orthogonal rotators, with the rotation axis tilted towards Earth. 
(J1705$-$1906 would be an orthogonal rotator with its rotation axis nearly perpendicular to the line of sight.)
For the naive assumptions of Figure \ref{AlphaZeta}, that is, ignoring effects that tend to align or unalign the magnetic
and rotation axes, and ignoring radio detectibility, of order 15\% of all pulsars would have such broad peaks, 
for this particular implementation of an OG model.
For the spin periods of PSRs J1055$-$6028 and J1913+0904, $\rho \approx 16^\circ$.
PSR J1913+0904's very low flux density of 70 $\mu$Jy at 1.4 GHz \citep{lfl+06} could indicate that only the edge
of its radio beam skims the Earth. The width of the radio profile of J1913+0904 is rather
narrow and the swing of the PPA is remarkably flat. 
This too could indicate that the line of sight cuts far from the magnetic axis.

Intuitively, pulsation tests work better when pulses are sharp.
To quantify this, we have simulated a range of pulse profile shapes, from Vela-like to ``box-like'', as for J1055$-$6028 and J1913+0904.
We add these signals to background, uniform in phase, and calculate the H-test for thousands of trials.
Fig. \ref{AlphaZeta}, right, shows how often the H-test significance exceeds $4\sigma$, as a function of signal strength.
(The background intensity is left fixed, and corresponds to the region surrounding Vela.) 
A broad box-like signal (width $w = 70$\%) is detected as often as a Vela-like signal only if it is $1.7$ times stronger.
In consequence we suspect that pulsars from that $(\alpha, \zeta)$ range are under-represented in the LAT sample.

The Atlas also predicts that aligned pulsars (small $\alpha$) viewed equatorially (large $\zeta$) will emit
gamma rays with essentially no phase modulation at all. 
Neither the H-test nor any other periodicity test will find these.
They would appear as steady sources with spectra typical of gamma-ray pulsars (see Eq. 1) at the positions of radio pulsars with large $\dot E/d^2$.
If too faint, spectral analysis will not reveal their similarity with gamma-ray pulsars.
Section 7 of 2PC partially addresses this topic, identifying 11 pulsars where the spectral shape in the off-peak phase intervals resemble those of gamma-ray pulsars.
Conversely, Table 13 of 2PC includes a few pulsars spatially coincident with steady gamma-ray sources, for which gamma-ray
pulsations have not been seen (gamma-quiet candidates).

The absence of strong off-pulse emission from PSR J1055$-$6028, apparent in Fig. \ref{WLCs_young}, bears mention.
Early in \textit{Fermi}'s mission, the LAT detected a flare from this direction \citep{2009ATel.2081}. 
\textit{Swift}/XRT observations (ObsIDs 31426, 31427) detected a faint X-ray source coincident with the luminous blue variable star AG Car (also
called WR 31b) suggesting it might be the counterpart \citep{2009ATel.2083}. 
Our subsequent analysis revealed the XRT counts were predominantly in the 0.3$-$2 keV band with rate $1.1 \pm 0.4$ counts ksec$^{-1}$ 
(observed flux $2.5 \times 10^{-14}$ erg cm$^{-2}$ s$^{-1}$ assuming a power-law with photon index 2.5). 
In fact, another potential counterpart to the LAT source is GG Car. 
Although not significantly detected in the \textit{Swift} observation, it is a binary system of two very massive stars, 
and interactions between their winds could in principle lead to gamma-ray emission \citep{FermiCWBsearch}.  
Because the LAT off-pulse intensity matches the expected diffuse background level around PSR J1055$-$6028 
we infer that none of the massive stars are bright gamma-ray sources, except perhaps episodically.


\section{Conclusions \& Prospects}
We see more atypical gamma-ray pulsars as \textit{Fermi}'s mission continues and fainter signals become detectable,
allowing us to probe a broader range of pulsar parameter space. 
In particular, we report the characteristics of six new gamma-ray pulsars which have in common that
they all have some property that delayed their discovery for a few years.

We have developed a method to ``diagnose'' the reliability of distances obtained using the NE2001 model and radio dispersion measures (DM),
where we compare the model's predictions for the electron density along the line of sight with other density tracers.
In the case of PSR J1055$-$6028 we conclude that the pulsar is at half the distance previously thought.
We also reported a timing parallax distance for PSR J1640+2224 using Nan\c cay data.

We have calculated the dependence of pulsed detection sensitivity on the pulse's duty cycle,
showing that this affects the uniformity with which we sample pulsars with different inclination angles.
Pulsed detection efficiency decreases but remains acceptable for very broad pulsations.

Data reconstructed with the ``\textit{Pass 8}'' analysis under development are being tested within the LAT instrument team \citep{Pass8}.
Pass 8's greatly increased acceptance at low energy changes which pulsars can rise above the diffuse emission; 
that is, it changes the pulsar detection biases in the LAT data and will help us acquire a more complete sample.


{\bf Acknowledgements}

The Nan\c cay Radio Observatory is operated by the Paris Observatory, associated with the French Centre National de la Recherche Scientifique (CNRS). 

The Parkes radio telescope is part of the Australia Telescope which is funded by the Commonwealth Government for operation as a National Facility managed by CSIRO. 
We thank our colleagues for their assistance with the radio timing observations.

The Lovell Telescope is owned and operated by the University of Manchester as part of the Jodrell Bank Centre for Astrophysics with 
support from the Science and Technology Facilities Council of the United Kingdom.

The \textit{Fermi} LAT Collaboration acknowledges generous ongoing support
from a number of agencies and institutes that have supported both the
development and the operation of the LAT as well as scientific data analysis.
These include the National Aeronautics and Space Administration and the
Department of Energy in the United States, the Commissariat \`a l'Energie Atomique
and the Centre National de la Recherche Scientifique / Institut National de Physique
Nucl\'eaire et de Physique des Particules in France, the Agenzia Spaziale Italiana
and the Istituto Nazionale di Fisica Nucleare in Italy, the Ministry of Education,
Culture, Sports, Science and Technology (MEXT), High Energy Accelerator Research
Organization (KEK) and Japan Aerospace Exploration Agency (JAXA) in Japan, and
the K.~A.~Wallenberg Foundation, the Swedish Research Council and the
Swedish National Space Board in Sweden.

Additional support for science analysis during the operations phase is gratefully
acknowledged from the Istituto Nazionale di Astrofisica in Italy and the Centre National d'\'Etudes Spatiales in France.

We thank Jonathan Braine for useful discussions about the HI and CO maps used to constrain the DM distances. 
C.C.C.~was supported in part by \textit{Swift} GI program 08-SWIFT508-0054.

\bibliographystyle{aa}
\bibliography{2ndPulsarCatalog}

\end{CJK}               
\end{document}